\documentclass[preprint,12pt]{elsarticle}

\usepackage{amssymb}

\newcounter{bla}

\journal{Computer Physics Communications}
\usepackage{graphicx,bm,amsmath,amssymb,xfrac,float,url}
\usepackage{xcolor}
\usepackage{subfig}
\usepackage{listings} 
\definecolor{lightgray}{gray}{0.95}
\usepackage[colorlinks=true,linkcolor=blue]{hyperref}
\usepackage{soul}
\setstcolor{red}

\begin{document}

\begin{frontmatter}



\title{CLAMM: a spin \textbf{CL}uster expansion--Monte Carlo toolkit for \textbf{A}lloys and \textbf{M}agnetic \textbf{M}aterials} 



\author[a]{Brian J. Blankenau\corref{author}}
\author[b]{Tianyu Su}
\author[a]{Namhoon Kim}
\author[a]{Elif Ertekin\corref{author}}

\cortext[author] {Corresponding author.\\\textit{E-mail address:}
ertekin@illinois.edu}
\address[a]{Department of Mechanical Science and Engineering, The Grainger College of Engineering, University of Illinois at Urbana-Champaign, Urbana, IL 61801.}
\address[b]{Department of Material Science and Engineering, The Grainger College of Engineering, University of Illinois at Urbana-Champaign, Urbana, IL 61801.}

\begin{abstract}

Finite-temperature magnetism gives rise to many phenomena in alloy materials, such as magnetic phase transformations, short or medium range order in magnetic alloys, spin waves, critical phenomena, and the magnetocaloric effect.
Lattice models, such as the Ising, Potts, cluster expansion, and magnetic cluster expansion models, are powerful tools for studying complex magnetic alloys and compounds. 
In this paper we introduce CLAMM, which is a new open source toolkit for developing custom lattice models from density functional theory (DFT) data sets. 
The toolkit is comprised of three main components. 
The first component is CLAMM\_Prep, a python tool that converts data sets consisting of the Vienna \textit{Ab-initio} Simulation Package (VASP) \cite{Kresse1993,Kresse1994,Kresse1996,Kresse1996a} DFT simulations into a compact format. 
The second component, CLAMM\_Fit, is also python-based and uses the compact data set to parameterize a lattice model, chosen from a set of available options (cluster expansion, Ising, and others).
The third component is CLAMM\_MC, which is a C++ Monte Carlo solver for generating ensembles of configurations, accounting for both magnetic and alloy configurational entropies, at different temperatures. 
These ensembles and their analysis can be used for simulating phase transformations and constructing phase diagrams. 
The code can also be used for generating special quasi-random structures and structures with user-defined short-range order. 
This document provides a comprehensive overview of each CLAMM tool in order to demonstrate CLAMM's potential for the computational materials community. 






\end{abstract}


\end{frontmatter}



{\bf PROGRAM SUMMARY}

\begin{small}
\noindent
{\em Program Title:} CLAMM -- A spin \textbf{CL}uster expansion--Monte Carlo toolkit for \textbf{A}lloys and \textbf{M}agnetic \textbf{M}aterials \\
{\em CPC Library link to program files:} (to be added by Technical Editor) \\
{\em Developer's repository link:} \url{https://github.com/ertekin-research-group/CLAMM} \\
{\em Code Ocean capsule:} (to be added by Technical Editor) \\
{\em Licensing provisions:} MIT License \\
{\em Programming language:} Python C++\\
{\em Nature of problem:} Lattice models, such as the Ising model and the cluster expansion method, are effective in capturing the thermodynamic properties of complex alloy systems.
However, constructing and validating such models can be complicated due to the need to accurately describe the temperature dependence of both configurational and magnetic disorder. \\
{\em Solution method:} CLAMM, a new lattice model toolkit designed for magnetic alloys, provides the flexibility for the user to specify different implementations of lattice or spin-lattice models with user-defined sublattices and order parameters.  \\
{\em Additional comments including restrictions and unusual features:} Required packages, installation, and tutorials can be found on the Github page. \\
   \\


\end{small}

\clearpage

\section{Introduction}

\begin{figure}
    \centering
    \includegraphics[width=1\linewidth]{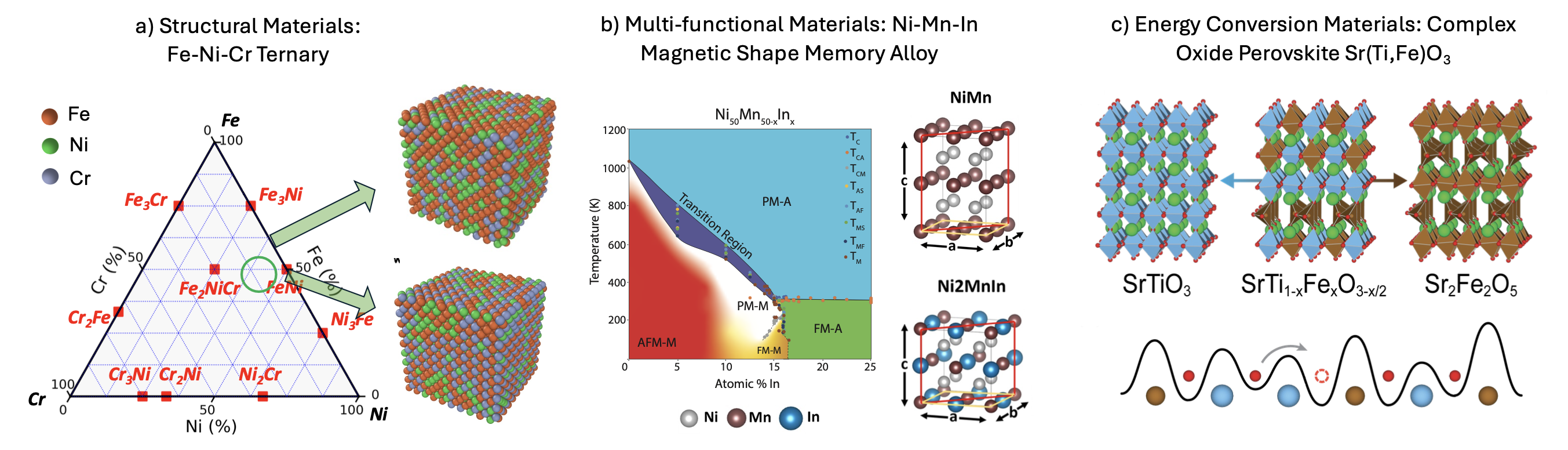}
    \caption{Magnetic alloys and compounds play critical roles in numerous applications from a) structural materials where magnetic interactions play an important role in hydrogen embrittlement of stainless steels \cite{Su2024First-principlesSteels}, to b) multi-functional materials such as the magnetic shape memory alloy Ni-Mn-In, to c) energy conversion materials such as Sr(Ti,Fe)O$_3$ where magnetism plays a key part in determining the short-range order and diffusivity \cite{Kim2019AtomicSr2Fe2O5}. }
    \label{fig:magalloy-ex}
\end{figure}

Lattice models such as the Ising model \cite{Ising1925BeitragFerromagnetismus,Onsager1944CrystalTransition}, the Potts model \cite{Potts1952SomeTransformations}, the cluster expansion  \cite{Sanchez1984GeneralizedSystems}, and the magnetic cluster expansion  \cite{Yu2015,VanDerVen2018First-PrinciplesCrystals,Puchala2013ThermodynamicsCalculations,Thomas2013Finite-temperaturePrinciples} are valuable tools for studying compounds, alloys, and magnetic systems. 
These methods provide a framework by which complex systems can be reduced to simpler, effective models while still capturing the critical physics. 
Such frameworks are useful across diverse technologically relevant materials, as shown in Figure \ref{fig:magalloy-ex}.  
For example in austenitic Fe–Ni–Cr alloys, structural properties, short-range order (SRO), and magnetism are inexorably linked \cite{Su2024First-principlesSteels} (Figure \ref{fig:magalloy-ex}(a)). 
Magnetic cluster expansion models are also useful for describing the complex magnetism of the austenite and martensite phases in magnetic shape memory alloys, such as the Ni-based Heusler alloys (Figure \ref{fig:magalloy-ex}(b)). 
Moreover, many perovskite oxide mixtures used in electrochemical systems for ion conduction (e.g., fuel and electrolysis cells) contain magnetic elements that influence defect association and dissociation reactions (Figure \ref{fig:magalloy-ex}(c)). 
Effective lattice models are well suited for these applications because they can capture the thermodynamic, temperature-dependent properties of magnetic alloys by accounting for finite temperature magnetism and chemical ordering/configuration. 
However, building and validating such models can be complicated due to the need to describe the effects of temperature on both configurational and magnetic disorder. 


In this work, we report a new lattice model toolkit designed for magnetic alloys named CLAMM (spin \textbf{CL}uster expansion--Monte Carlo toolkit for \textbf{A}lloys and \textbf{M}agnetic \textbf{M}aterials).
CLAMM allows the user to specify different implementations of lattice or spin-lattice models with user-defined sublattices and order parameters. 
CLAMM consists of three toolkits: (i) data preparation, (ii) model fitting, and (iii) Monte Carlo simulations. 

\textit{Relationship to existing tools and codes:} Because of the success and popularity of lattice models, many lattice model codes and tools have been developed and are widely available within the computational materials community. 
One of the most widely used cluster expansion codes is the \textit{Alloy Theoretic Automated Toolkit}, or ATAT, developed by van de Walle et al. \cite{VandeWall2002Self-drivenDiagrams,VandeWalle2002TheGuide,VanDerVen2018First-PrinciplesCrystals}. 
ATAT is actively maintained at the time of this writing.
ATAT possesses functionality for generating \textit{ab initio} based phase diagrams and special quasi-random structures (SQS), but currently lacks the capability to incorporate spin models to account for magnetism. 
CASM, recently released in 2021, is another powerful lattice model code \cite{Thomas2013Finite-temperaturePrinciples,Puchala2013ThermodynamicsCalculations,VanDerVen2018First-PrinciplesCrystals,CASM}. 
It was developed by van der Ven et al. \cite{CASM} and is explicitly designed to implement magnetic cluster expansions for the purpose of studying complex magnetic compounds and alloys. 
To our knowledge, it is the only other general purpose code that explicitly includes the capability to apply lattice models to magnetic alloys. 
Compared to these existing tools, a main distinguishing aspect of CLAMM is the utilization of the ``decorated cluster expansion'', or DCE, as developed by \cite{Kim2019AtomicSr2Fe2O5}, in which each cluster is uniquely described by a motif and decoration rather than  traditional cluster functions, as described later. 
The use of the DCE offers both interpretable models and compatibility with Ising-like spin models. 
Other distinguishing aspects of CLAMM 
include: the ability to include order parameters for tracking separate magnetic states (A, C, and G anti-ferromagnetism), the ability to handle multiple user-defined sublattices (i.e. a multi-sublattice cluster expansion), and a general flexibility to include specific user-designated clusters such as clusters dependent on local composition. 
CLAMM also has the capability to find all symmetry equivalent clusters based on the point group symmetry analysis given the cluster and the parent lattice. 
This capability is enabled by the symmetry package based on the spglib \cite{Togo2024Spglib:Search}, as implemented in the Python Materials Genomics (pymatgen) Python library \cite{Ong2013PythonAnalysis}.

\section{Workflow and Overview}
\label{Section_WorkflowOverview}

\subsection{Basic Workflow}

CLAMM consists of three interconnected toolkits: CLAMM\_Prep, CLAMM\_Fit, and CLAMM\_MC. 
\begin{figure}
    \centering
    \includegraphics[width=5in]{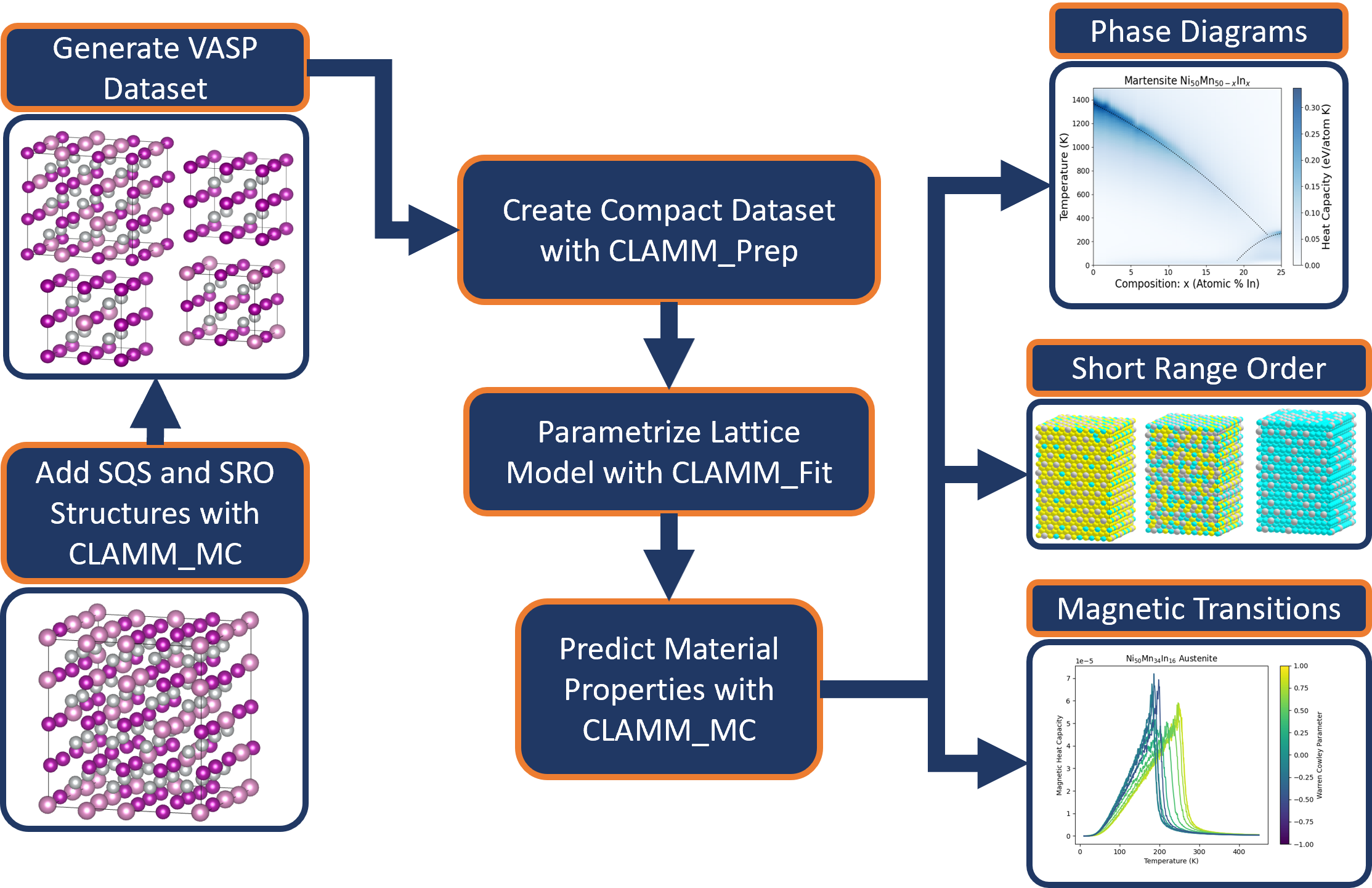}
    \caption{ CLAMM workflow: The user supplies a data set of DFT simulations, some of which may include POSCAR files generated with CLAMM\_MC. The data set is compiled into a compact format with CLAMM\_Prep which in turn is used by CLAMM\_Fit to parameterize a lattice model. Then CLAMM\_MC then uses the parameterized model to calculate materials properties such as magnetic transitions and phase diagrams.}
    \label{fig:workflow}
\end{figure}
CLAMM\_Prep is a tool used to analyze outputs of DFT simulations produced with the Vienna Ab initio Simulation Package \cite{Kresse1993,Kresse1994,Kresse1996} and collates the outputs to a compact format. This compact data set is used by CLAMM\_Fit to determine a user-selected lattice model's cluster contribution terms (CCTs), which are the fitted parameters of the model. 
These CCTs could be the effective cluster interaction terms (ECI) for a cluster expansion, the magnetic exchange constants of a spin-lattice model, or any user-defined model parameter. 
Finally, CLAMM\_MC is a Monte Carlo solver that can be used to generate SQS, simulate phase transformations, analyze the effects of short-range order, or predict an alloy's compositional stability. An example workflow utilizing the CLAMM toolkit is shown in Figure \ref{fig:workflow}. 
This workflow will be discussed in detail in this paper.

\subsection{Data Set Generation}

With the exception of structure generation, such as generating SQS structures or structures with user-defined degrees of short-range order, the CLAMM workflow begins with building a data set of DFT simulations. 
For the purpose of constructing a lattice model, the initial data set must contain a diversity of atomic and magnetic configurations, and the configurations that make up the data set must span the intended composition space.
While CLAMM does not currently have the functionality to automatically generate a DFT data set, it provides a tool to assist in this process. 
This tool is designed to make the data set generation process more efficient by generating certain VASP \cite{Kresse1993,Kresse1994,Kresse1996,Kresse1996a} input such as POSCAR files and corresponding magnetic configurations. 
When building effective models for complex magnetic alloys, manually generating configuration files can quickly become cumbersome given the large amount of data required. 
CLAMM's short-range order optimization algorithm enables users to automatically generate new configurations with the user-prescribed degree of short-range order, provided that the systems are large enough to accommodate the statistical nature of any Monte Carlo solver. 
This process will be discussed in more detail in section 2.5. 
By contrast, for small ordered systems (around 16 atoms or less), it is still necessary to create the atomic configurations (POSCAR files) and magnetic initialization (MAGMOM tag in INCAR files) manually. 

After all necessary input files are created, VASP simulations must be carried out, as CLAMM pulls results from CONTCAR and OUTCAR files. 
The results of all DFT simulations must be organized in individual folders under a common home directory. 
The contents of each folder for a given DFT simulation must contain a valid POSCAR, CONTCAR, and OUTCAR file. Should any of these files be missing, corrupt, or incomplete, the user will be notified when executing the relevant CLAMM code.

As described below, when processing the DFT results, a user-defined crystal structure with placement of atoms on well-defined lattice sites is assumed. 
Data processing requires extracting information from relaxed DFT structures about the distribution of atoms on expected sites and identification of first, second, etc nearest neighbors.
Typically, lattice models (cluster expansion, Ising) are agnostic to inter-atomic distances, regardless of changes in composition or configuration, since they predict total energies entirely based on atomic coordination environments. 
Parameters such as bond-distances are often only implicitly included in lattice models, to the extent that local environment correlates to bond lengths. 
In the process of generating the DFT data set to obtain configuration energies however, it is often necessary to perform full relaxations of lattice shape, atomic positions, and lattice volume which can adjust the atoms' positions from their initial, high symmetry configuration. 
For the purposes of CLAMM, DFT calculations can be allowed to fully relax as the atomic positions and lattice constants are taken from the original POSCAR file in order to identify atomic site occupations, neighbor distributions, and other features associated with the expected crystal structure. 
However, limits should be placed on the magnitude of these relaxations to ensure that both files represent the same phase (the relaxation should not alter the crystal structure or move the phase to a different symmetry that is not expected to be described by the lattice model). 
The lattice constants and atomic positions are read from both the POSCAR and CONTCAR files, and a tolerance is applied to their deviation from the expected positions. 
Should this tolerance be exceeded, the DFT simulation will not be included in the data set. 
Any model parameterized with CLAMM should be tested with different deviation tolerances to ensure model convergence and stability. 
Another important consideration when generating a data set for CLAMM is the shapes of the included structures. 
In CLAMM, it is not necessary that all systems in the data set are supercells of the same underlying unit cell, only that they can be mapped to the same primitive cell with minimal variation of inter-atomic distances. 
For example, a data set may include a structure represented by the primitive cell as well as by a conventional cell so long as all inter-atomic distances and coordination are identical. 

\subsection{Preparing data sets with CLAMM -- CLAMM\_Prep}

Following data set construction and execution of DFT simulations, the next step in the CLAMM workflow is data processing and re-formatting. 
The necessary tools for this step are contained in the CLAMM\_Prep Python code. 
CLAMM\_Prep is a relatively simple code with three main functionalities: collating the results of DFT simulations into a single data file, identifying corrupt or incomplete simulation results, and formatting the simulation results to be consistent with a lattice model representation. 
The data collation and filtering happen automatically upon running the CLAMM\_Prep python script, while the formatting requires user input in the form of a param\_in file. An example param\_in file for a Ni$_2$MnIn data set is shown in Listing \ref{lst:PreproIn}.
\begin{lstlisting}[caption= Example param\_in file for CLAMM\_Prep.,label={lst:PreproIn}]
root_dir: '/Users/Ni2MnIn_data set'
do_compile_datad: True
write_data: 'data_mag'
species: ['Ni', 'Mn', 'In'] 
lat_tol: [0.1, 0.1, 0.1]
pos_tol: [0.01, 0.01, 0.01]
read_mag: True
use_spin_tol: True                
spin_tol: [[0.2], [3.0, 2.0], [-1]] 
do_mag_distrib: True
write_mag_distrib: 'mag_distrib'  
\end{lstlisting}
Line 1 sets a home directory to search through for DFT simulations, line 2 states whether or not to create a new data set file, and line 3 sets the output file name and location for the collated data set. 
All atomic species that may appear in the data set are listed in line four. 
In this example, the data set would consist of structures with a maximum of three atomic species. 
The order in which the species are listed should be consistent with the order in which they are listed in the VASP POSCAR files. 
The lat\_tol and pos\_tol flags describe the allowed deviation between POSCAR and CONTCAR files in terms of $a$, $b$, and $c$ lattice constants and $x$, $y$, and $z$  atomic coordinates, respectively. 

For cases where magnetism is included in the desired model, the read\_mag must be set to True, in which case total magnetic moments on each atomic species will be read from the OUTCAR file. 
In this case, it is necessary to assign spin values to each atom in each configuration, to be used in lattice models. 
The assigned spin values are based on the atom's final magnetic moment at the end of the DFT simulation.
There are two methods for formatting information on magnetic moments in the final data set: using the raw magnetic moment, or setting a moment cutoff. 
The desired method is set using the use\_spin\_tol flag (line 8). 
When the flag is set to ``True'', the spin tolerance method is used. 
Spin values are typically assigned integer values such as ${-1,0,1}$ or ${-1,1}$. 
It is necessary to define a map from the atom's final magnetic moment in the DFT simulation to an allowed set of integers.
In the example in Listing \ref{lst:PreproIn}, the cutoff option is selected.  
On line 9, magnetic moment cutoffs are listed for each atom in the same order as in line 4. 
For the first atom, in this case Ni, there are two allowable spin magnitudes:
\begin{itemize}
    \item The magnitude of the DFT magnetic moment is greater than 0.2, resulting in an assigned spin with a magnitude of 1, and a sign that matches the sign of the DFT magnetic moment.
    \item The magnitude of the magnetic moment is less than 0.2, resulting in a spin with a magnitude of 0.
\end{itemize}
For the second atom (Mn, in the second set of brackets) there are three allowable spin magnitudes:
\begin{itemize}
    \item The magnitude of the magnetic moment is greater than 3.0, resulting in an assigned spin with a magnitude of 2, and a sign that matches the sign of the DFT magnetic moment.
    \item The magnitude of the magnetic moment is less than 3.0 but greater than 2.0, resulting in a spin with a magnitude of 1, and a sign that matches the sign of the DFT magnetic moment.
    \item The magnitude of the magnetic moment is less than 2.0, resulting in a spin with a magnitude of 0. 
\end{itemize} 
A tolerance of 0 indicates all spins for that element are set to zero, as shown in the example above for In. 
This is a reasonable option for elements that are expected to be non-magnetic.
Alternatively, in order to designate an element as having only one spin magnitude uniformly applied across all configurations (such as $\|$spin$\|$ = 1, spin = sgn(DFT atomic moment), the tolerance is set to 0. 

\begin{figure}
    \centering
    \includegraphics[width=3in]{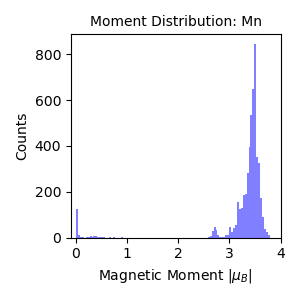}
    \caption{A histogram showing the distribution of magnitudes of magnetic moments for Mn in a sample dataset.}
    \label{fig:mom_dist}
\end{figure}

Determining the cutoff values for assigning spins is critical for developing a spin-lattice model. 
We caution that users should test models using different tolerances, assess and evaluate differences, and be aware of the possible sensitivity of their results to parameter selection. 
To help make this determination, CLAMM\_Prep can output a list of magnetic moments for each element in the data set. 
This list is accomplished by setting the do\_mag\_distrib flag (line 10 in Listing \ref{lst:PreproIn} to True and supplying a name and location of a magnetic moment list file (line 11 in Listing \ref{lst:PreproIn}). 
The magnetic moment list can then be used to create a histogram displaying the distribution of magnetic moments for each species, contained in the whole data set. 
An example histogram for the distribution of Mn moments for the Heusler alloy Ni$_{50}$Mn$_{50-x}$In$_x$ data set is shown in Figure \ref{fig:mom_dist}. 
The histogram shows three groups with different magnetic moments, which are indicated at 0 $\mu_B$, 2.8 $\mu_B$, and 3.5 $\mu_B$. 
To illustrate the possible complexities that may arise, due to the comparatively low occurrence of moments around 2.8 $\mu_B$, it is possible that it is of minimal significance and that this group should be included with the larger group centered at 3.5 $\mu_B$. 
Alternatively, the group could indicate the existence of a distinct but infrequently observed magnetic moment for the elemental species, which should be treated differently in the lattice spin model.
This can be determined by comparing the resulting models using different moment cutoffs.

After running the CLAMM\_Prep script, an output file will be generated containing an entry for every structure in the data set. 
An example of a data set consisting of one Ni$_2$MnIn DFT simulation is shown below in Listing \ref{lst:out}.
The first line lists the elements present in the structure in the same order as mentioned above. 
The second line contains the number of each individual element in the configuration, the name of the directory where the DFT simulation was found, the relaxed energy of the structure in eV, and the lattice constants with lattice angles given in radians. 
The following three lines contain the lattice vectors taken from the initial POSCAR file. 
The remaining lines contain the index, atomic species (enumerated in the order shown in line one), assigned spin, and fractional coordinates of each atom. 
With the data set file now created, it can be used by CLAMM\_Fit to create and parameterize a model. 

\begin{lstlisting}[caption= Example CLAMM\_Prep output file.,label=lst:out]
# Ni Mn In
8 4 4 \Ni2MnIn\Mart\B0 -88.558 5.2 5.2 5.2 1.571 1.571 1.571
5.2 0  0
0  5.2  0
0  0  7.8
   0   0   1.0   0.25   0.25   0.25
   1   0   1.0   0.75   0.25   0.25
   2   0   1.0   0.25   0.75   0.25
   3   0   1.0   0.75   0.75   0.25
   4   0   1.0   0.25   0.25   0.75
   5   0   1.0   0.75   0.25   0.75
   6   0   1.0   0.25   0.75   0.75
   7   0   1.0   0.75   0.75   0.75
   8   1   1.0   0.5    0.0    0.0
   9   1   1.0   0.0    0.5    0.0
   10  1   1.0   0.5    0.0    0.5
   11  1   1.0   0.0    0.5    0.5
   12  2   0     0.0    0.0    0.0
   13  2   0     0.5    0.5    0.0
   14  2   0     0.0    0.0    0.5
   15  2   0     0.5    0.5    0.5
\end{lstlisting}

\subsection{Model Parametrization -- CLAMM\_Fit}

The CLAMM\_Fit code is used to fit the free parameters of a desired lattice model using the compact data set from CLAMM\_Prep and a model definition file consisting of cluster motifs and cluster types.
Specifically, an ordinary linear system of equations that can be expressed by 
\begin{equation}
\begin{bmatrix} N_{11} & N_{12} & ... & N_{1M} \\ N_{21} & N_{22} & ... & N_{2M} \\
... \\ 
N_{N1} & N_{N2} & ... & N_{NM}
\end{bmatrix} 
\left\{ \begin{array}{c} J_1 \\ J_2 \\ ... \\ J_M \end{array} \right\} = 
\left\{ \begin{array}{c} E_1 \\ E_2 \\ ... \\ E_N \end{array} \right\} 
\end{equation}
is formulated, in which $E_i$ is the total energy of configuration $i$ from DFT, $N_{ij}$ is the number of times that motif and type $j$ appears in configuration $i$, and the CCTs $J_j$ are the free parameters (magnetic or cluster expansion) to be fitted. 
Here, $i = 1 \, ... \, N$ denotes the configuration number, and $j = 1 \, ... \, M$ denotes the motif and type. 

With the cluster motifs and types specified in the model definition file, CLAMM\_Fit first evaluates each configuration in the compact data set to identify $N_{ij}$, the number of occurrences of each motif and type $i$ within each configuration $j$. 
The CCTs for each cluster $J_j$ can be determined by performing a linear regression on the occurrences from each DFT simulation and their corresponding energies.
The final model as described by each cluster and their corresponding CCTs is then output to a file named CLUSTERS.  
The file can later be used by the Monte Carlo solver CLAMM\_MC to build an effective Hamiltonian to use for subsequent Monte Carlo simulations.

\begin{figure}
    \centering
    \includegraphics[width=3in]{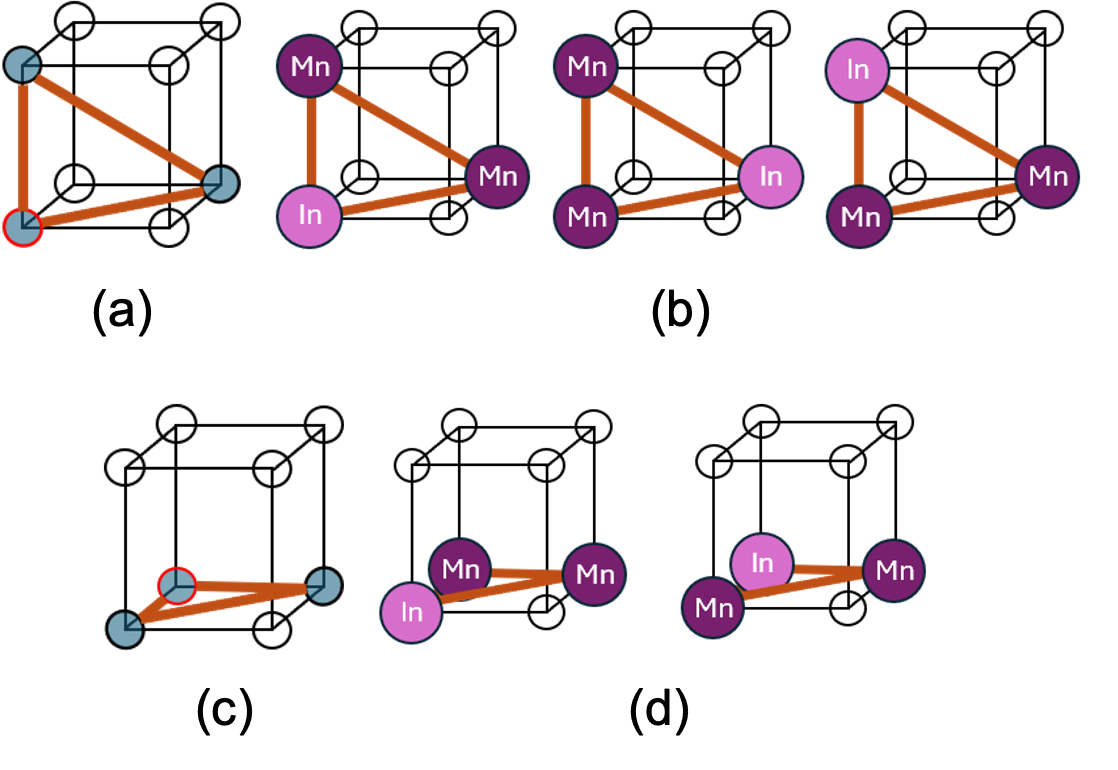}
    \caption{Cluster definitions in the DCE framework. a) An example of a DCE motif. b) The symmetrically unique clusters for motif a) with a decoration of two Mn and one In atoms. c) A DCE motif with higher symmetry. d) The symmetrically unique clusters for motif c) with a decoration of two Mn and one In atoms.}
    \label{fig:clust_motif}
\end{figure}

CLAMM has the flexibility to incorporate a number of different lattice models. 
Currently, cluster expansions, $n$-spin Ising models,  $n$-state Potts models, and linear combinations of these models are supported. 
Unlike most cluster expansion codes, CLAMM implements the representation of the cluster expansion model as proposed by Kim et al. \cite{Kim2019AtomicSr2Fe2O5}, in which each cluster is uniquely described by a motif and decoration rather than by using traditional  cluster functions \cite{Sanchez1984GeneralizedSystems,vandeWalle2002AutomatingCalculations}. 
The Kim interpretation of the cluster expansion will be referred to as the decorated cluster expansion or DCE. 
An example of a cluster in the DCE is depicted in Figure \ref{fig:clust_motif}. 
Separate motifs are shown in Figure \ref{fig:clust_motif}(a,c) along with examples of unique decorations of two Mn atoms and one In atom (Figure \ref{fig:clust_motif}(b,d)). 
For the motif in Figure \ref{fig:clust_motif}a, there are three symmetrically unique decorations for the case of two Mn and one In atoms. 
Each of these unique decorations, in combination with the motif, defines its own cluster. 
A detailed comparison of the conventional and DCE implementations can be found in Ref.~\cite{KIM2022110969}. 
An advantage of the DCE over the conventional cluster expansion is the ease with which the clusters identified can be interpreted. 
However, an additional advantage for the purpose of CLAMM is the ease with which DCE clusters and spin-lattice clusters can be represented in the same user-supplied input file, by supplying only the cluster type and motif.
Listing \ref{lst:motif} defines four motifs and two clusters. 
The first motif (line two) describes the set of monomer clusters for a cluster expansion (type 0). 
The second motif (line three) describe the set of all two-atom clusters with a neighbor distance of 2.6 \r{A} for a cluster expansion. 
The third motif describes the set of all three atom clusters having relative positions of [0, 0, 0], [2.6, 0.0, 0.0], [2.6, 0.0, 0.0], again for a cluster expansion model. 
The final motif (line 5) describes the set of all two-atom clusters with neighbor distance of 2.6 \r{A} for an Ising model.

\begin{figure}
    \centering
    \includegraphics[width=3in]{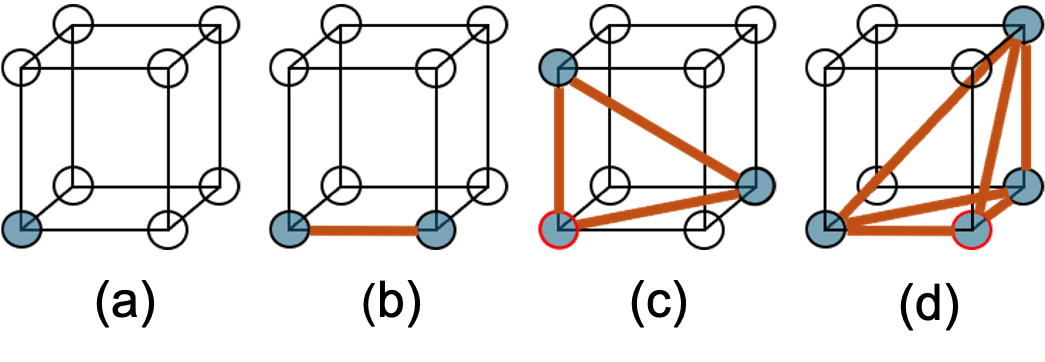}
    \caption{Four example cluster motifs. a) An example of a monomer motif. b) An example of a dimer motif. c) an example of a trimer motif. d) An example of a tetramer motif}
    \label{fig:clusts}
\end{figure}

\begin{lstlisting}[caption= Example motif definition file param2\_in for Figure \ref{fig:clusts}.,label=lst:motif]
{"List": [
[[0,0,0],[1],[0]],
[[[0,0,0],[2.6, 0, 0]],[1],[1]],
[[[0,0,0],[2.6, 2.6, 0],[0, 0, 3.9]],[1],[0]],
[[[0,0,0],[-2.6,0,0],[0,2.6,0],[0.0,2.6,3.9]],[1],[0]],
]}
\end{lstlisting}
An example of a simple model description file, param2\_in, is shown above in Listing \ref{lst:motif} and the motifs listed in the file are depicted in Figure \ref{fig:clusts}.
The input file has a JSON readable format, with each line representing a new cluster motif. 
The atomic positions of each constituent atom are listed in the first set of brackets. 
In line two of Listing \ref{lst:motif} the motif describes a monomer and therefore the atomic positions are [0, 0, 0]. The motif described in line three is a dimer with atomic positions [[0, 0, 0], [2.6, 0, 0]]. 
In each motif description, position is given with respect to a reference atom ([0, 0, 0]). 
By convention, the reference atom is chosen to be one with the smallest neighbor distances. 
To avoid any ambiguity in this convention, the reference atom is defined as the atomic site which minimizes
\begin{equation}
    d_i = \sum_{j\neq i} | r_i-r_j |      \hspace{0.5em},
\end{equation}
where the sum is over all atomic sites in a given motif and $r_i$ is the position of a reference atom. 
The atomic positions can be given in absolute or fractional coordinates. 
However, if fractional coordinates are used, a scaling factor must be supplied. 
The scaling factor follows immediately after the atomic positions in each motif description. 
In the above example, positions are given in absolute coordinates, so the scaling factor is set to 1. 
For each input cluster, the point group symmetry operations are found using the pymatgen symmetry package \cite{Ong2013PythonAnalysis}. 
These symmetry operations are then applied during the counting process to obtain all clusters that are symmetry-equivalent to the input cluster. 
By mapping the symmetry-equivalent clusters on the DFT structures/configurations, all possible atomic decorations of the target cluster can be found and the counts of the motifs can be calculated. 

The final entry in each motif definition is the cluster type. 
For the example motif file in Listing \ref{lst:motif}, the motif in line two has a type of [0], and the motif in line three has a type of [1].
The cluster type defines the type of lattice model applied to each cluster of a given motif type. The currently supported lattice models are:
\begin{itemize}
    \item type 0: cluster expansion using DCE representation
    \item type 1: $n$-spin Ising model ($n=$ number of discrete spin states) 
    \item type 2: Three body $n$-spin Ising model (Developmental)
    \item type 3: $n$-state Potts model ($n=$number of discrete spin states). 
\end{itemize}
For each motif type, the appropriate model is applied when tabulating the occurrences of each cluster in a DFT configuration. 
For example, the cluster occurrence for cluster $\beta$ in an $n$-spin Ising model is given by 
\begin{equation}
    Y(\beta) = \sum_{\alpha \in A}  \delta_{\alpha,\beta}\sigma^{\alpha}_1 \sigma^\alpha_2       \hspace{0.5em} 
\end{equation}
where the sum is over the set of all clusters $\alpha$ present in a given DFT simulation $A$. 
For every cluster $\alpha$ equivalent to $\beta$, the cluster occurrence $Y$ is incremented by the product of the spins for the two sites in the $\alpha$ cluster, $\sigma^\alpha_1$ and $\sigma^\alpha_2$.
In the case of a cluster expansion, the occurrence of cluster $\beta$ is given by:
\begin{equation}
    Y(\beta) = \sum_{\alpha \in A} \delta_{\alpha,\beta}       \hspace{0.5em}.
\end{equation}
In this case the occurrence is incremented for each cluster $\alpha$ equivalent to $\beta$. 
Similarly, the expressions for the $n$-state Potts model are given by

\begin{equation}
Y(\beta) = \sum_{\alpha \in A} \delta_{\alpha,\beta}\delta_{\sigma_1, \sigma_2}\hspace{0.5em}.
\end{equation}

Depending on the set of clusters included in the model and the magnetic and atomic configurations present in the data set, it is possible that not all configurations can be described by a unique set of cluster occurrences. 
That is, it is possible that two different configurations have the same cluster distribution (the same number of each type of cluster), when decomposed.  
The simplest example of how this could occur is if only one-body (monomer) terms are included in a cluster expansion for binary alloy AB. In this case any configuration that is a 50-50 alloy will have the same cluster decomposition.
When this cluster degeneracy occurs, only the structure with the lowest DFT energy is kept for model-fitting purposes. 
The directories where the discarded structures can be found are printed along with a warning, but the program will continue to run. 

With a complete set of cluster occurrences for each DFT configuration, it is then possible to parameterize the model (solve for the CCTs) using linear regression. 
The user can select from four separate linear regression methods: LASSO, Ridge, Least Squares, and Elastic Net. 
The relevant regularization parameters are determined using cross-validation. 
For each method, bootstrapping is implemented to provide an estimate of error for each model parameter. 
The resulting set of model parameters is recorded in an output file called CLUSTERS. 
An example output for the clusters defined in Listing \ref{lst:motif} is shown in Listing \ref{lst:clustExOut}.


The effective Hamiltonian for the compound lattice model defined in Listing \ref{lst:clustExOut} takes the form:
\begin{equation}
\label{eqH}
H(\vec{\theta},\vec{\sigma})=\sum_{\alpha \in A} J_\alpha\Theta_\alpha(\vec{\theta}) + \sum_{\beta \in B} J_\beta\Omega_\beta(\vec{\sigma}) + \sum_i^N \mu_i h \sigma_i  \hspace{0.5em},
\end{equation}
\begin{equation}
    \Omega_\beta(\vec{\sigma}) = \sum_{\beta^{'} \in \beta} \sigma_{1}^{\beta^{'}}\sigma_{2}^{\beta^{'}}    \hspace{0.5em}.
    \label{eqO}
\end{equation}
Here, the first sum in Equation (\ref{eqH}) is the cluster expansion in the DCE representation (type 0), and the second and third sums are the Ising model in the DCE-like representation (type 1). 
The vector $\vec{\theta}$ is the decoration of all atomic sites, and $\vec{\sigma}$ is the spin of all atomic sites.  
Together, $\vec{\theta}$ and $\vec{\sigma}$ describe the configuration of a given system.
The first summation in Equation (\ref{eqH}) is over the set of all clusters $A$ of the cluster expansion.
For an individual cluster $\alpha$, the CCT (here, an effective cluster interaction) is given by $J_\alpha$, and the occurrence of $\alpha$ in a configuration is given by $\Theta_\alpha(\vec{\theta})$. 
The second summation is over the set of clusters $B$ for the Ising model. 
The constant for cluster $\beta$ is given by CCT $J_\beta$ (here, an exchange interaction). 
The occurrence for an Ising cluster is given by Equation (\ref{eqO}), where the summation is over all symmetry-equivalent clusters $\beta$ and $\sigma_1^{\beta^{'}}$ and $\sigma_2^{\beta^{'}}$ are the spins on the atoms in cluster $\beta^{'}$. The final summation is over all atoms in a given configuration and captures the effects of an external magnetic field of strength $h$, and species-specific magnetic moment $\mu_i$. 

In addition to the model definition file, two additional input files are needed to run the CLAMM\_Fit code. 
The first is a structure file, formatted as a VASP POSCAR file.  
This file is used for purposes of identifying coordination between all atomic sites defining lattice symmetries, which aid in the tabulation of cluster occurrences. 
The second is a param\_in file, which is used to configure fitting options. 
An example param\_in file for CLAMM\_Fit is shown in Listing \ref{lst:fitparam}. 
\begin{lstlisting}[caption= An example param\_in file for CLAMM\_Fit, label=lst:fitparam]
do_count: False                          
do_fit: True                          
lat_in: 'POSCAR'                      
data_file: 'Mart_pp'                  
clust_in: 'cluster_in'                
species: ['Ni', 'Mn', 'In']           
fit_ridge: True                      
rescale_enrg: False                   
sample_times: 1000                  
sample_ratio: 0.9                  
kfold: 10                        
alpha_range: [-4, 2]             
convergence: 1e-5                
\end{lstlisting}

For very large data sets and for models with a large number of clusters, especially three-body terms and higher, the tabulation of cluster occurrences is a lengthy process. 
Therefore, it is sometimes preferable to break up the occurrence tabulation and model fitting into separate tasks. 
This is accomplished with the DO\_COUNT and DO\_FIT flags. 
If the DO\_COUNT flag is set to True, cluster occurrences will be tabulated for all DFT simulations in the data set. 
Upon completion, all necessary information to perform a linear regression for a given data set will be written to a data\_save file. 
If the DO\_COUNT flag is set to False, then Gen\_MC will check to see if a data\_save file exists. 
If such a file is found, then all relevant information will be read from the file, and CLAMM\_Fit will proceed with model parameterization, provided that the DO\_FIT flag is set to true. 
Lines 3, 4, and 5 in Listing \ref{lst:fitparam} provide the name and location of the input POSCAR file, the data set file, and the model definition file, respectively. 
Line 6 lists the atomic species in the data set, which are enumerated to define all possible cluster decorations. 
In line 7, the linear regression method used to parameterize the model here was set to Ridge regression. 
The re-scale energy flag has been set to false. 
This flag determines whether or not the regression should fit the total energy per atom or re-scale to the energy above the convex hull. 
The sample\_times and sample\_ratio flags are optional, and used for bootstrapping purposes and define the number of bootstrapped data sets to be used and the fraction of DFT simulations in the full data set to be sampled for each bootstrapped set. 
Finally, the $k$-fold alpha range and convergence settings are used to determine the regularization parameter for the regression method via cross-validation.
All fitting routines use the implementations available in the python scikit-learn library.

\begin{lstlisting}[caption= Parametrized model file (CLUSTERS) for motifs defined in Listing \ref{lst:motif}, label=lst:clustExOut]
Motif : intercept 
Enrg : -6.298548712827071
#
Type : 0
Motif :
0.0, 0.0, 0.0
Deco : 0 : 1 : 2
Enrg : 0.0 : 0.0 : 0.001231
#
Type : 0
Motif :
0.0, 0.0, 0.0 : 2.6, 0.0, 0.0
-2.6, 0.0, 0.0 : 0.0, 0.0, 0.0
0.0, 0.0, 0.0 : 0.0, 2.6, 0.0
0.0, -2.6, 0.0 : 0.0, 0.0, 0.0
Deco : 2, 2 : 1, 2 : 2, 1 : 1, 1 : 0, 0
Enrg : 0.117 : 0.042 : 0.042 : -0.045 : 0.0
#
Type : 0
Motif :
0.0, 0.0, 0.0 : 0.0, 2.6, 0.0 : 2.6, 0.0, 0.0
0.0, 0.0, 0.0 : -2.6, 0.0, 0.0 : 0.0, -2.6, 0.0
0.0, 0.0, 0.0 : -2.6, 0.0, 0.0 : 0.0, 2.6, 0.0
0.0, 0.0, 0.0 : 0.0, -2.6, 0.0 : 2.6, 0.0, 0.0
-2.6, 0.0, 0.0 : -2.6, 2.6, 0.0 : 0.0, 0.0, 0.0
2.6, 0.0, 0.0 : 0.0, 0.0, 0.0 : 2.6, -2.6, 0.0
0.0, -2.6, 0.0 : -2.6, -2.6, 0.0 : 0.0, 0.0, 0.0
0.0, 2.6, 0.0 : 0.0, 0.0, 0.0 : 2.6, 2.6, 0.0
-2.6, 0.0, 0.0 : -2.6, -2.6, 0.0 : 0.0, 0.0, 0.0
2.6, 0.0, 0.0 : 0.0, 0.0, 0.0 : 2.6, 2.6, 0.0
0.0, 2.6, 0.0 : -2.6, 2.6, 0.0 : 0.0, 0.0, 0.0
0.0, -2.6, 0.0 : 0.0, 0.0, 0.0 : 2.6, -2.6, 0.0
Deco : 2,2,2:2,1,2:2,2,1:2,1,1:1,2,2:1,1,2:1,2,1:1,1,1:0,0,0
Enrg : 0.302:0.022:0.022:0.032:0.031:0.011:0.011:-0.016:0.0
#
Type : 1
Motif :
0.0, 0.0, 0.0 : 2.6, 0.0, 0.0
-2.6, 0.0, 0.0 : 0.0, 0.0, 0.0
0.0, 0.0, 0.0 : 0.0, 2.6, 0.0
0.0, -2.6, 0.0 : 0.0, 0.0, 0.0
Deco : 2, 2 :1, 2: 2, 1: 1, 1: 0, 0
Enrg : 0.0: 0.0: 0.0: 0.039: -0.001
#
\end{lstlisting}

With all input files supplied, the CLAMM\_Fit code can be run and the resulting parameterized model will be written in the format shown in Listing \ref{lst:clustExOut} to file CLUSTERS.
The fitted model parameters are organized by motif in the order supplied in Listing \ref{lst:motif}, with each motif delineated by a $\#$ and all CCT energies given in eV per cluster instance. 
All CCT energies in this example are purely place holders and are not the result of an actual model fit. 
The first entry represents the model intercept and all subsequent entries represent the motifs defined in Listing \ref{lst:clustExOut}. The first motif, beginning on line 5, is relatively straightforward, as it  represents a monomer cluster. First, the cluster type is given, followed by the motif definition, the motif decorations, and finally the CCTs associated with each combination of motif and decoration. 
The formatting of the motif in the third entry, beginning on line 11, requires additional explanation. 
This motif, as with all motifs consisting of more than one atom, is represented by separate lines for all symmetrically equivalent translations and rotations of the original motif from Listing \ref{lst:clustExOut}. 
This format is optimized for subsequent Monte Carlo simulations using CLAMM\_MC,  
as listing symmetrically equivalent motifs explicitly in this form enables more efficient algorithms in the Monte Carlo implementation.

\subsection{Applying Lattice Models with CLAMM -- CLAMM\_MC}

With CLAMM\_MC, a fully parameterized model can be applied in conjunction with Monte Carlo sampling, to calculate statistical properties or generate new configurations under varied thermodynamic or compositional conditions. 
CLAMM\_MC is written in C++ and optimized to run on a Linux operating system. 
In order to run CLAMM\_MC, a minimum of three input files must be supplied. The three primary input files are a POSCAR file, a INPUT file, and a parameterized model definition file such as the CLAMM\_Fit output file discussed in the previous section. 
Additional input files may be required depending on the use case. 

The POSCAR file serves as the unit cell for CLAMM\_MC and provides the basic geometry as well as sublattice geometry for subsequent simulations. The POSCAR file needed for CLAMM\_MC is compatible with but varies slightly from the typical POSCAR file used in VASP. In addition to the lattice vectors and atomic positions listed in a POSCAR file, CLAMM\_MC also requires definitions of specific sublattices. 
This is achieved by stipulating the allowed atomic species at each site in the lattice. 
An example POSCAR for a Ni$_2$MnIn lattice is shown below in Listing \ref{lst:POS}. 
The allowed atomic species for a given atomic site are listed following the XYZ positions of each atom. 
\begin{lstlisting}[caption= A sample input POSCAR file for CLAMM\_MC, label=lst:POS]
Ni2MnIn
6.01
    1   0   0
    0   1   0
    0   0   1
Ni Mn In
8  4  4
Direct
0.25  0.25  0.75 Ni
0.25  0.75  0.75 Ni
0.75  0.25  0.75 Ni
0.75  0.75  0.75 Ni
0.25  0.25  0.75 Ni
0.25  0.75  0.75 Ni
0.75  0.25  0.75 Ni
0.75  0.75  0.75 Ni
0.0   0.0   0.0  Mn In
0.0   0.5   0.0  Mn In
0.5   0.0   0.0  Mn In
0.5   0.5   0.0  Mn In
0.0   0.0   0.5  Mn In
0.0   0.5   0.5  Mn In
0.5   0.0   0.5  Mn In
0.5   0.5   0.5  Mn In
\end{lstlisting}
This example describes two sublattices: one for Ni and one for Mn and In. 
During Monte Carlo simulations, Mn and In atoms could be placed or swapped with any atomic site on their sublattice but never on the Ni sublattice. 
From the supplied POSCAR file, CLAMM-MA can generate a simulation supercell of arbitrary dimensions. However, if a specific configuration, or specific initial configuration is required, it is recommended to include the full desired simulation cell in the POSCAR file.

The INPUT file for CLAMM\_MC defines the type of simulation to be run, as well as several relevant simulation parameters. 
An example INPUT file is shown in Listing \ref{lst:input}. 
The ALGO option is used to select the CLAMM\_MC functionality that is required for the desired simulation. The current implemented ALGO options are:
\begin{itemize}
    \item ALGO = -2: An implementation of a Metropolis Mote Carlo algorithm with options for simulated annealing and support for a developmental three-atom-cluster Ising model. 
    \item ALGO = -1: An implementation of a Metropolis Monte Carlo algorithm optimized for generating atomic configurations with user-defined short-range order.
    \item ALGO = 0: A model evaluation option. The effective Hamiltonian supplied by the parameterized model is evaluated for a given POSCAR file or derived supercell along with any magnetic order parameters. This is useful for debugging purposes. 
    \item ALGO = 1: An implementation of a Metropolis Monte Carlo algorithm with options for simulated annealing designed for systems where only magnetic degrees of freedom are allowed to change.
    \item ALGO = 2: An implementation of a Metropolis Mote Carlo algorithm with options for simulated annealing designed for systems where only the atomic configuration is allowed to change.
    \item ALGO = 3: An implementation of a Metropolis Mote Carlo algorithm with options for simulated annealing where both atomic and magnetic configurations are allowed to change.
    \item ALGO = 4: An implementation of a Metropolis Mote Carlo algorithm with options for simulated annealing where both atomic and magnetic configurations are allowed to change. This option includes support for systems with vacancies and interstitial defects.
\end{itemize}
\begin{lstlisting}[caption= Smaple INPUT file for CLAMM\_MC, label= lst:input]
ALGO = 3
STRUCTURE = POSCAR
USE_POSCAR = TRUE
SHAPE = 1 1 1
ATOM_NUMBS = 1728 864 864
SPECIES = Ni Mn In
SIM_TYPE = DEFAULT
SPIN_INIT = FM
TA_PASSES = 100
EQ_PASSES = 300
START_TEMP = 4000
END_TEMP = 0
TEMP_STEP = 40
SRO_TARGET = 1.0
USE_STATES = TRUE
WRITE_CONTCARS = FALSE
\end{lstlisting}
The SHAPE option, line 4, defines the supercell dimensions of the simulation cell, and the ATOM\_NUMBS setting gives the number of atoms for each atomic species. 
If the numbers for ATOM\_NUMBS are inconsistent with the proportions of atomic species present in the POSCAR file, then the CLAMM\_MC will attempt to distribute atoms in the simulation cell to be consistent with ATOM\_NUMBS. 
The SPIN\_INIT option is used to set the initial magnetic order. The current options are FM, AFM, and RAND for random initialization. 
Additional options for user-defined magnetic configurations are under construction and have not yet been fully implemented. 
The settings in lines 9 through 13 set the parameters for the Metropolis Monte Carlo implementation and for simulated annealing. 
TA\_PASSES refers to equilibration time in Monte Carlo passes before thermal averaging of measured parameters begins. 
EQ\_PASSES refers to the number of Monte Carlo passes in a thermal average. 
TEMP\_STEP refers to the temperature increments to be taken during simulated annealing between START\_TEMP and END\_TEMP. SRO\_TARGET is an optional parameter used when generating a structure with user-defined short-range order (ALGO = -1). 
The short-range order parameter used when this parameter is set is the Warren-Cowley parameter \cite{Warren1941X-RayLattices,Cowley1950AnAlloys}. 

The SRO\_TARGET option on line 14 sets the target value for the Warren-Cowley parameter.
If this option is used, an SRO definition file must also be provided. 
This file has a format almost identical to the model definition file; however, it contains only a single motif. 
The line starting with ``Enrg'' indicates the weight of the SRO parameters which the code optimizes (and in particular does not refer to the CCT energy; the name is retained for consistency with the cluster files). 
An example SRO definition file is shown in Listing \ref{lst:SROdef}. 

\begin{lstlisting}[caption= Example SRO definition file for CLAMM\_MC. This can also serve as a spin-product definition file by changing the Type option from 0 to 1, label=lst:SROdef]
Type : 0
Motif :
0.0, 0.0, 0.0 : 2.6, 0.0, 0.0
-2.6, 0.0, 0.0 : 0.0, 0.0, 0.0
0.0, 0.0, 0.0 : 0.0, 2.6, 0.0
0.0, -2.6, 0.0 : 0.0, 0.0, 0.0
Deco : 2, 2 : 1, 2 : 2, 1 : 1, 1 : 0, 0
Enrg : 0 : 1 : 1 : 0 : 0
#
\end{lstlisting}
The SRO definition file can also serve to define a spin-product magnetic order parameter. 
In a typical spin lattice model, it is common to use order parameters to quantify the magnetic configuration. 
The most common order parameter is the average magnetization ($M$) given by
\begin{equation}
   M(\vec{\sigma})=\frac{1}{N}\sum^N_i \sigma_i \hspace{0.5em},
\end{equation}
where the sum is over all $N$ atomic sites $i$, $\sigma_i$ is the spin at site $i$ and $\vec{\sigma}$ is a given magnetic configuration. This parameter is useful in distinguishing between FM and PM states. 
However, $M$ is insufficient to distinguish between antiferromagnetic and paramagnetic states, since in both cases the total magnetic moments are expected to sum to zero. 
Using a spin-product magnetic order parameter, it is possible to distinguish between multiple types of magnetic configurations with zero net magnetization such as alternate types of antiferromagnetism and paramagnetism.
The spin-product magnetic order parameter is defined by:
\begin{equation}
       M^{'}(\vec{\sigma_\alpha})=\frac{1}{N_\alpha}\sum^N_{i_\alpha,j_\alpha} \sigma_{i_\alpha}\sigma_{j_\alpha} \hspace{0.5em}.
\end{equation}
Here the summation is over all $N_\alpha$ pairs of atomic sites $i_\alpha$ and $j_\alpha$ which belong to a two-atom cluster $\alpha$. 
The neighbor distance considered for the spin product can be defined in the SRO definition file. 

Referring again to the CLAMM\_MC INPUT file in Listing \ref{lst:input}, the USE\_STATES tag is used to define the magnitude of allowed spins for each individual atomic species. 
These allowed spins are read from the SPIN\_STATES file. 
An example SPIN\_STATES states file is shown in Listing \ref{lst:spinstates}.  
\begin{lstlisting}[caption= A SPIN\_STATES file used to supply the allowed spins for each Ni\, Mn\, and In atom, label = lst:spinstates]
-1 0 1
-1 0 1
0
\end{lstlisting}
Here, each line lists the allowed spins on each atom in the order they appear in the INPUT and POSCAR files. In this example, Ni and Mn atoms are allowed spins of $(0,\pm1)$ and all In atoms have their spins set to zero.
The final flag in the example CLAMM\_MC input file is the WRITE\_CONTCARS flag. If the WRITE\_CONTCARS flag is set to True, then after all Monte Carlo passes have been completed at a given temperature, current state of the simulation (the decoration and spin for each atomic site) is written using a modified POSCAR format. In this modified format, the atomic species and spin of each atom are listed following the atom’s XYZ coordinates. If WRITE\_CONTCARS is set to False, only the final simulation state (the decoration and spin for each atomic site following the final Monte Carlo move) will be printed.

In addition to the CONTCAR files, an additional output file named OUTPUT is always generated. 
This file contains the thermodynamic or order parameters for each temperature step. 
The specific contents of this file vary depending on the ALGO flag selected. 
However, the OUTPUT file for each ALGO flag follows a common format. 
An example is shown in Listing \ref{lst:output}
\begin{lstlisting}[caption= A sample OUTPUT file used to supply the allowed, label = lst:output] 
Using Algo3
Composition: 1728 864 864 
MC passes: 300
Initial spin energy per atom: -2.36347e-05
Initial spin per atom: -0.0179398
T, E, M, var_e, var_s, Cmag, Xmag, flip_count, flip_count2
2000, -3.3486e-05, -0.00110, 0.02446, 0.26510, 7.09748e-05, 7.69109, 769620, 916952
1999, -4.44865e-05, -0.00049, 0.02726, 0.26888, 7.9169e-05, 7.80848, 771486, 915168
\end{lstlisting} 
Lines one through six provide information on the simulation initialization. Lines one and two and three display the name of the algorithm, the simulation composition, and the number of thermal averaging steps selected in the INPUT file. Lines four and five show the initial energy and initial net magnetization for the system before any Monte Carlo steps are preformed. The remaining lines contain the thermal averaged data for each temperature step. In the case of the example in listing \ref{lst:output} this includes the energy, net magnetization, energy variance, spin variance, magnetic heat capacity, magnetic susceptibility, and total number of accepted and rejected Monte Carlo moves.

\begin{figure}
    \centering
    \includegraphics[width=0.65\linewidth]{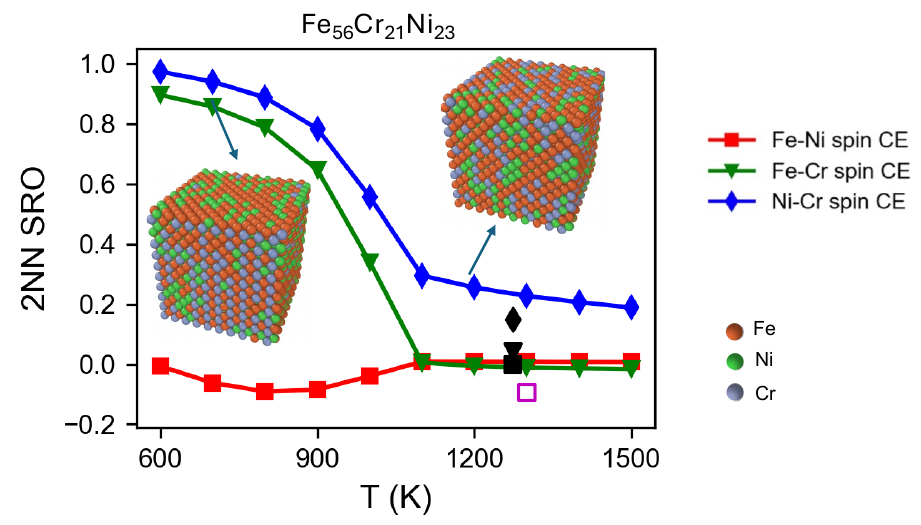}
    \caption{Warren-Cowley short-range order in Fe-Ni-Cr alloys as a function of temperature. Blue, red, and green data points were generated from CLAMM\_MC simulations \cite{Su2024First-principlesSteels}. Black data points were taken from experiment\cite{Cenedese1984DiffuseFe0.56Cr0.21Ni0.23}. }
    \label{fig:TianyuMC}
\end{figure}

Beyond the primary results contained in the OUTPUT file, CLAMM\_MC can provide other output options as well. 
One such option is used for the temperature-dependent Warren-Cowley short-range order parameter, which is often used to study short-range atomic arrangements at different temperatures.
This data is available when setting ALGO to 2 or 4. In this can, an additional output file will be produced named OUTPUT\_SRO. This file contains the frequency $P(\sigma)$ with which each cluster $\sigma$ appeared during a given temperature step. For two-atom clusters, the Warren-Cowley parameter described in terms cluster $\sigma$ $\alpha$ can then be calculated as:
\begin{equation}
\alpha_{\sigma} = 1-\frac{P_{\sigma}}{2\theta_a\theta_b}  \hspace{0.5em},
\end{equation}
where $\theta_a$ and $\theta_b$ are the concentrations of the atomic species that decorate $\sigma$. Figure \ref{fig:TianyuMC} illustrates an application of CLAMM\_MC for studying short-range order in Fe-Cr-Ni alloys. 
Using the CLAMM-MA tool-set, Su et al. showed that Cr concentration and magnetic structure are key factors is determining the short-range order of Fe-Cr-Ni alloys \cite{Su2024First-principlesSteels}. 
In Figure \ref{fig:TianyuMC} it is clear that, as the temperature decreases, Cr-based clusters promote ordering while Fe-Ni clusters remain randomly distributed.

\subsection{Monte Carlo Implementation} 

\begin{figure}
    \centering
    \includegraphics[width=5.25in]{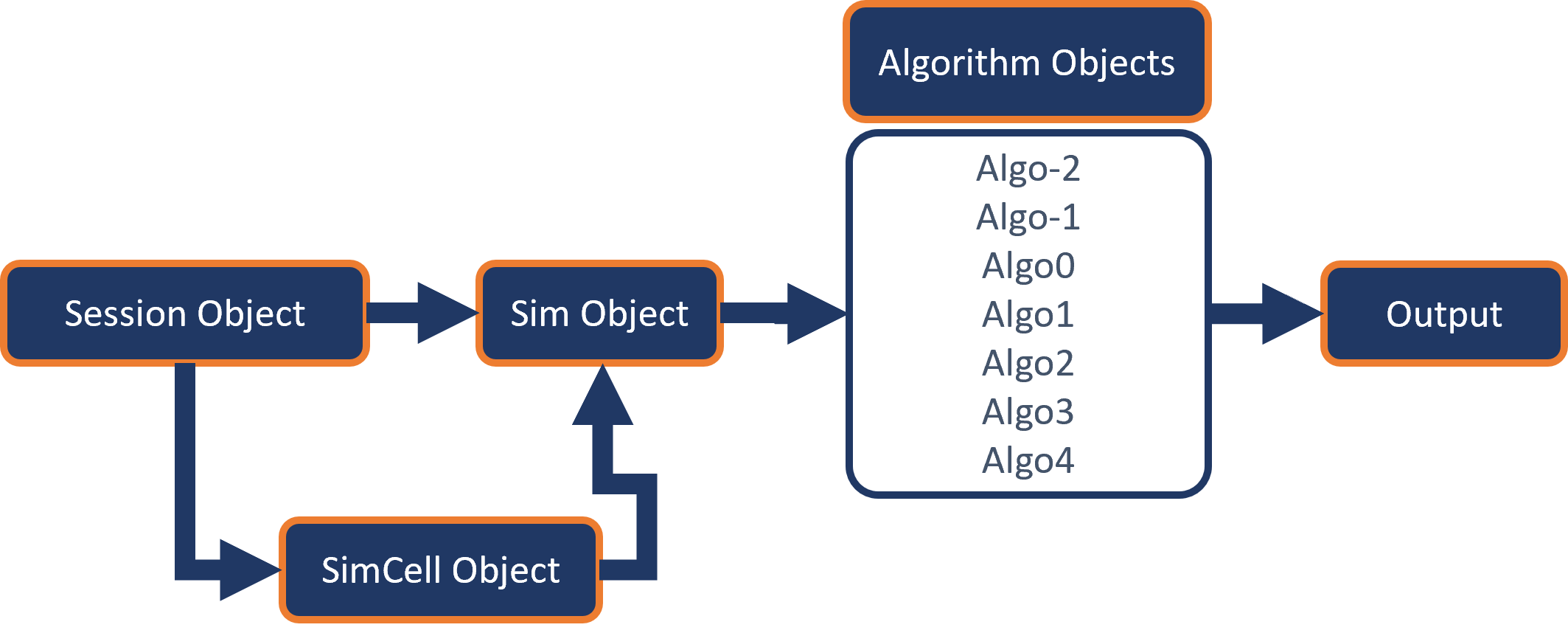}
    \caption{Object relationships for CLAMM\_MC. The Session object contains the user input from the INPUT, lattice model, SRO, and $M'$ definition files. It is used along with the POSCAR file to create the SimCell object. These are then passed to the appropriate algorithm object through the Sim object.}
    \label{fig:CLAMM-MA_run}
\end{figure}

Part of the motivation behind the development of CLAMM\_MC was to create an environment flexible enough to effectively handle different complex multi-sublattice magnetic alloys. 
In keeping with this goal, CLAMM-MA was developed with a modular design that is capable of being tailored to and optimized for different types of lattice models and Monte Carlo algorithms. 
At the heart of this modularity are four C++ objects, the Session object, the SimCell object, the Sim object, and the algorithm object. 
A diagram illustrating the relationships between these objects is found in Figure \ref{fig:CLAMM-MA_run}.
The Session object contains copies of information from the input files so that the Monte Carlo simulation carried out can be documented for reference. 
This includes the settings and flags from the INPUT file and SPIN\_STATES file, the lattice model provided by the parameterized model file (CLUSTERS), and any short-range order or spin-product definitions found in the input files directory. 
The Session object is also used to create the SimCell object.
The SimCell object contains an initial copy of the simulation cell, constructed based on the settings found in the Session object, the lattice, and the sublattices defined in the POSCAR file.  
The SimCell object uses these inputs to populate its primary member variable, the atom list. 
The atom list is a vector of atom objects. 
The atom object contains the current atomic species, site specific allowed atomic species and corresponding allowed spins, atomic position, and a neighbor list. The neighbor list is created to include only atoms allowed by the lattice model definition.  
After the Session and SimCell objects are created, it would, in principle, be possible to operate on these two objects to perform any desired calculation or simulation. 
However, these objects and the data structures they contain are not intended for this purpose. 
Instead, they are intended to serve as a translation layer between the user input and the algorithm objects, which are optimized for specific tasks and simulations. 
The Session and SimCell objects are passed to the desired algorithm object through the Sim object. The code defining the Sim object is shown in Listing \ref{lst:Sim}. 
\begin{lstlisting}[caption= The Sim object C++ code, label=lst:Sim]
#include "run_sim.h"

Sim::Sim(void) {}

Sim::Sim(Session& _session, SimCell& _sim_cell) {
	sim_cell = _sim_cell;
	session = _session;
}

void Sim::start() {
    cout << "Using algo " << session.algo << "\n";
    if (session.algo == -1) { return; }
    else if (session.algo == -2) { Algo_m2 algo(session, sim_cell); algo.run(); }
    else if (session.algo == 0) { Algo0 algo(session, sim_cell); algo.run(); }
    else if (session.algo == 1) { Algo1 algo(session, sim_cell); algo.run(); }
    else if (session.algo == 2) { Algo2 algo(session, sim_cell); algo.run(); }
    else if (session.algo == 3) { Algo3 algo(session, sim_cell); algo.run(); }
    else if (session.algo == 4) { Algo4 algo(session, sim_cell); algo.run(); }
}
\end{lstlisting}
The Sim object contains only a single function, start(). 
This function references the Session object to determine the desired algorithm to be run, creates an instance of that particular algorithm object, passes the Session and SimCell objects to the algorithm objects as arguments, and then executes the function algorithm.run(). 
If a user desires functionality that is not covered by the currently implemented algorithm objects and wishes to contribute their own, they must add it as an option in the Sim.start() function by assigning it to a currently unused integer. 
For example, if a new algorithm was developed named test\_algorithm they would add the following line after line 18 in Listing \ref{lst:Sim}. 
\begin{lstlisting}[caption= Adding a ALGO option to CLAMM\_MC, label=lst:Sim_2]
        else if (session.algo == 5) { test_algorithm algo(session, sim_cell); algo.run(); }
\end{lstlisting}
They must also include the header file of their new algorithm class in the Sim header file.
We have found that this approach, in which a dedicated, problem-specific version of the desired Monte Carlo simulation is compiled before execution, results in faster execution times than a one-size-fits-all approach. 

\subsection{Note for Developers}
The constructer for each algorithm class must take as arguments a reference to the Session and SimCell objects. 
Additionally, every algorithm class must contain the function run().  
Apart from these requirements any new algorithm class may be developed completely independently of the rest of the CLAMM-MA project. 
As mentioned in the previous section, there are currently seven implemented algorithm classes. 
These algorithms represent many of the most common calculations and simulations used for magnetic alloys and compounds. 
When developing a new algorithm class, it is not necessary to follow the same format as these already implemented classes. 
However, it may be useful to any future developers to understand the methods and formats used in the current algorithms.  
Each of these algorithms have been implemented to be at least partially optimized for speed. 
When dealing with any large simulation or any large multi-body simulation, one is forced to consider the trade-off between memory requirements and algorithm execution speed.  
In the case of the currently implemented algorithms, we have chosen to favor speed over memory concerns. 
The limiting factor in the speed of these algorithms has proven to be the evaluation of the lattice model for a given atomic site or group of atomic sites. 
To construct a more efficient method of evaluating the lattice model, the Simcell and Session objects were used to create a compact description for the state of the simulation at any point during a simulation run. 
Each cluster is reformatted so that it can be represented as a vector of integers. 
The first integer is a unique index, and the remaining integers encode the motif and decoration. 
The vectors are then grouped according to the type of cluster they represent (eg. Ising or cluster expansion). 
These vectors are then used to create a hash table according to the hash function shown in Listing ref{lst:hash}, which links each hash with the appropriate  CCT. 
\begin{lstlisting}[caption = Hash function for the cluster vector, label=lst:hash]
size_t Algo4::cust_hash(vector<uint32_t>& vect) {
    std::size_t seed = vect.size();
    for (auto x : vect) {
        x = ((x >> 16) ^ x) * 0x45d9f3b;
        x = ((x >> 16) ^ x) * 0x45d9f3b;
        x = (x >> 16) ^ x;
        seed ^= x + 0x9e3779b9 + (seed << 6) + (seed >> 2);
    }
    return seed;
}
\end{lstlisting}
This hash function is not guaranteed to be free of collisions for any arbitrary vector, however we have tested for collisions with the set of vectors that are possible with up to 600 clusters, including up to four body terms and consisting of up to four separate atomic species. 
We have found that no collisions exist within this limit.  We cannot guarantee that larger systems will be free of hash collisions.
The Simcell object is used to create a similar vector representing each atomic site as well as a separate vector containing each cluster index which can be applied to the atomic site. 
To evaluate the lattice model at a given atomic site, these two vectors are combined and sequentially passed to the hash function. 
The resulting CCTs are then combined in accordance with the cluster type. In principle, this procedure could be made even more efficient by replacing the encoding of the clusters as vectors with bitmaps. 
While the current implementation allows for a relatively efficient evaluation of the lattice model it does first require the creation of the necessary vectors and tables. 
As a result, for small simulations, up to half of the total execution time of a simulation may be taken up by creating the necessary data structures.

\section{Conclusion} 

In this work we have discussed the CLAMM-MA toolset for lattice model tools to describe the thermodynamic properties of complex magnetic alloys. 
Such alloys underlie a large number of emerging technologies, such as magnetic shape memory alloys, structural materials and medium/high entropy alloys, and electrochemical oxide mixtures used in batteries and in fuel/electrolysis cells. 
This toolset is divided into three packages CLAMM\_Prep, CLAMM\_Fit, and CLAMM\_MC. 
CLAMM\_Prep is used to construct compact representations of data sets composed of DFT simulations carried out with the VASP software package. 
CLAMM\_Fit provides a flexible framework for defining and parameterize lattice models for complex multi sublattice magnetic systems. 
The lattice models parameterized with CLAMM\_Fit are then used in CLAMM\_MC to predict materials properties, generate magnetic phase diagrams, evaluate phase stability, and generate special quasi random structures or magnetic structures with user-defined short-range order. 
A number of commonly used Monte Carlo algorithms have been implemented in CLAMM-MA. 
However, users and developers are able to contribute new algorithms to expand its capabilities, as a result of the modular construction of CLAMM-MA.
The software is available on Github under the MIT License: \url{https://github.com/ertekin-research-group/CLAMM}. 
The actively developed source code, instructions for installing the software, and tutorials can be found at this link. 
We encourage that CLAMM be used to help design and study complex magnetic alloys.

\section*{Author Contributions}

B.J.B., T.S., N.K., and E.E. conceived and designed this research project. B.J.B., T.S., and N.K. developed the software. E.E. supervised the project. All authors participated in preparing and editing the manuscript.

\section*{Conflicts of Interests}

There are no conflicts of interest to declare.

\section*{Acknowledgements} 

This work was supported by the US Department of Energy/National Nuclear Security Administration through the Chicago/DOE Alliance Center, cooperative agreements DE-NA0003975 and DE-NA0004153. 
This work used PSC Bridges-2 at the Pittsburgh Supercomputing Center through allocation MAT220011 from the Advanced Cyberinfrastructure Coordination Ecosystem: Services \& Support (ACCESS) program, which is supported by National Science Foundation grants \#2138259, \#2138286, \#2138307, \#2137603, and \#2138296.





\bibliographystyle{elsarticle-num}
\bibliography{bib}







\end{document}